\documentclass[12pt]{article}

\usepackage{amsmath,amsfonts,latexsym,amssymb,amscd}
\usepackage{pslatex}
\usepackage[latin1]{inputenc}
\usepackage[T1]{fontenc}
\usepackage{pspicture}
\usepackage{verbatim,amsthm,curves,graphics}
\usepackage{mathrsfs}

\usepackage{graphicx}

\newcommand{\mr}{{\mathbb R}}

\begin{document}


\title{ Area-angular momentum-charge inequality for  stable marginally outer
trapped surfaces in 4D Einstein-Maxwell-dilaton theory }

\author{
Stoytcho Yazadjiev$^{}$\thanks{\tt yazad@phys.uni-sofia.bg}
\\ \\
{\it $ $Department of Theoretical Physics, Faculty of Physics, Sofia
University} \\
{\it 5 J. Bourchier Blvd., Sofia 1164, Bulgaria} \\
    }
\date{}

\maketitle

\begin{abstract}
We derive inequalities between the area, the angular momentum and
the charges for axisymmetric  closed outermost stably marginally
outer trapped surfaces, embedded in dynamical  and, in general,
nonaxisymmetric spacetimes satisfying the
Einstein-Maxwell-dilaton-matter equations. In proving  the
inequalities we assume that  the dilaton potential is non-negative
and that the matter energy-momentum tensor satisfies the dominant
energy condition.
\end{abstract}


\sloppy

\section{Introduction}

 Dynamical black holes are a serious challenge to the present-day
investigations in general relativity and  alternative theories of
gravity. Black hole dynamics  is  very difficult to study within the
framework of the existing theoretical scheme,
 and consequently our understanding of
 dynamical black holes is not so deep as for  isolated
stationary black holes. In this situation, the derivation of certain
estimates and inequalities on the physical characteristics of
dynamical black holes, based mainly on "first principles" and
independent of the specific features of the dynamical processes, is
very  important. Within the general theory of relativity, lower
bounds for the area of dynamical  horizons in terms of their angular
momentum or/and charge were given in \cite{ADC}--\cite{CJR},
generalizing the similar inequalities for  stationary black holes
\cite{AHC}--\cite{HCA}. These remarkable inequalities are based
solely on general assumptions and they hold for any axisymmetric but
otherwise highly dynamical horizon in general relativity. For a nice
review on the subject we refer the reader to \cite{D}. The
relationship between the proofs of the area-angular-momentum-charge
inequalities for quasilocal black holes and stationary black holes
is discussed in \cite{CENS}-\cite{J}.

A natural problem is to find similar inequalities in alternative
theories of gravity which generalize Einstein theory. An example of
such a theory  is the so-called Einstein-Maxwell-dilaton gravity,
which naturally arises in the context the generalized scalar-tensor
theories of gravity,  the low energy string theory \cite{Gibbons,
Garfinkle}, and Kaluza-Klein theory \cite{Maison}, as well as  in
some theories with gradient spacetime torsion \cite{Hojman}.

The field equations of Einstein-Maxwell-dilaton gravity with matter
are presented below in eq. (\ref{EMDFE}). A characteristic feature
of this theory is the coupling between the scalar field (dilaton)
$\varphi$ and the electromagnetic field $F_{ab}$, and  this coupling
is governed by a parameter $\gamma$ (called dilaton coupling
parameter). The static and stationary isolated black holes in 4D
Einstein-Maxwell-dilaton theory were extensively studied in various
aspects during the last two decades. The classification of the
isolated stationary, axisymmetric, asymptotically flat black holes
with a connected horizon in Einstein-Maxwell-dilaton gravity was
given in \cite{SY} for dilaton coupling parameter $\gamma$
satisfying $0\le \gamma^2 \le 3$. The static asymptotically flat
Einstein-Maxwell-dilaton black holes (without axial symmetry and the
horizon connectedness assumption) were classified in \cite{UAMS}.
The sector of stationary Einstein-Maxwell-dilaton black holes with a
dilaton coupling parameter beyond the critical value $\gamma^2 = 3$
is extremely difficult to analyze analytically. Most probably the
black hole uniqueness is violated in this sector as the numerical
investigations imply \cite{Kunz}.

In the present paper we derive some inequalities between the area,
the angular momentum and the charges for dynamical black holes in
Einstein-Maxwell-dilaton gravity with a non-negative dilaton
potential and with a matter energy-momentum tensor satisfying the
dominant energy condition.

\section{Basic notions and setting  the problem }

Let $({\cal M},g)$ be a $4$-dimensional spacetime satisfying the
Einstein-Maxwell-dilaton-matter  equations
\begin{eqnarray}\label{EMDFE}
&&R_{ab} - \frac{1}{2}R g_{ab}= 2\nabla_a\varphi \nabla_b\varphi - g_{ab}\nabla^{c}\varphi\nabla_{c}\varphi
+ 2e^{-2\gamma\,\varphi} \left(F_{ac}F_{b}{}^c- \frac{g_{ab}}{4}F_{cd}F^{cd}\right) \nonumber \\ && - 2V(\varphi) g_{ab} + 8\pi T_{ab},\nonumber\\
&& \nabla_{[a} F_{bc]}=0, \\
&&\nabla_{a}\left(e^{-2\gamma\,\varphi}F^{ab}\right)=4\pi J^{b}, \nonumber\\
&&\nabla_{a}\nabla^{a}\varphi= -\frac{\gamma}{2}e^{-2\gamma
\,\varphi}F_{ab}F^{ab} +\frac{dV(\varphi)}{d\varphi}, \nonumber
\end{eqnarray}
where $g_{ab}$ is the spacetime metric,  $\nabla_{a}$ is its
Levi-Civita connection and $G_{ab}=R_{ab} - \frac{1}{2}R g_{ab}$  is
the Einstein tensor. $F_{ab}$ is the Maxwell tensor and $J^{a}$ is
the current.  The dilaton field is denoted by $\varphi$,
$V(\varphi)$ is its potential and $\gamma$ is the dilaton coupling
parameter governing the coupling strength of the dilaton to the
electromagnetic field. The matter energy-momentum tensor is
$T_{ab}$. We assume that $T_{ab}$ satisfies the dominant energy
condition. Concerning the dilaton potential, we assume that it is
non-negative ($V(\varphi)\ge 0$).

Further we consider a closed orientable 2-dimensional spacelike
surface ${\cal B}$ smoothly embedded in the spacetime ${\cal M}$.
The induced metric on ${\cal B}$ and its Levi-Civita connection are
denoted by $q_{ab}$ and $D_a$. respectively. In order to describe
the extrinsic geometry of ${\cal B}$ we introduce the normal
outgoing and ingoing null vectors $l^{a}$ and $k^{a}$ with the
normalization condition $g(l,k)=l^{a}k_{a}=-1$. The extrinsic
geometry then is characterized by the expansion $\Theta^{l}$, the
shear $\sigma^{l}_{ab}$, and the normal fundamental form
$\Omega^{l}_a$ associated with the outgoing null normal $l^{a}$ and
defined as follows:
\begin{eqnarray}
&&\Theta^{l}= q^{ab}\nabla_{a}l_{b}, \\
&&\sigma^{l}_{ab}= q_{a}^{c}q_{b}^{d} \nabla_{c}l_{d} -
\frac{1}{2}\Theta^{l}q_{ab}, \\
&&\Omega^{l}_{a} = - k^{c}q_{a}^{d}\nabla_{d}l_{c}.
\end{eqnarray}

In what  follows we require  ${\cal B}$ to be a marginally outer
trapped surface (i.e., $\Theta^{l}=0$) and ${\cal B}$ to be stable
(or spacetime stably outermost in more formal language)
\cite{H}--\cite{AMS1},\cite{JRD}. The last condition means that
there exists an outgoing vector $V^a=\lambda_1l^a - \lambda_2k^a$
with functions $\lambda_1\ge 0$ and  $\lambda_2>0$ such that
$\delta_{\,V} \Theta^l\ge0$, with $\delta_{\,V}$ being the
deformation operator on ${\cal B}$ \cite{AMS}--\cite{BF}. In simple
words the deformation operator describes the infinitesimal
variations of the geometrical objects on ${\cal B}$ under an
infinitesimal deformation of ${\cal B}$ along the flow of the vector
$V^a$.

As an additional technical assumption, we require ${\cal B}$ to be
invariant under the action of the $U(1)$ group with a Killing
generator $\eta^a$. We assume that the Killing vector $\eta^a$  is
normalized to have  orbits with a period $2\pi$. Also we require
that ${\cal B}$ is axisymmetrically stable\footnote{In other words,
axisymmetric and stable with axisymmetric functions $\lambda_1$ and
$\lambda_2$. }, $\pounds_\eta l^{a}=\pounds_\eta k^{a}=0$, and
$\pounds_\eta \Omega^{l}_a=\pounds_\eta {\tilde F}_{ab}=\pounds_\eta
\varphi=0$, where ${\tilde F}$ is the projection of the Maxwell
2-form on ${\cal B}$.

From the axisymmetric stability condition one can derive the
following important inequality valid for every axisymmetric function
$\alpha$ on ${\cal B}$ \cite{JRD}

\begin{eqnarray}
\int_{{\cal B}} \left[|D\alpha|^2_q  + \frac{1}{2}R_{{\cal
B}}\alpha^2\right]dS \ge \int_{\cal B}\left[\alpha^2
|\Omega^{\eta}|^2_q
 + \alpha\beta |\sigma^{l}|^2_q + G_{ab}\alpha l^{a}(\alpha
k^{b}+ \beta l^{b})\right] dS ,
\end{eqnarray}
where $|\,.\,|_{q}$ is the norm with respect to the induced metric
$q_{ab}$, $dS$ is the surface element measure on ${\cal B}$,
$R_{\cal B}$ is the scalar curvature of ${\cal B}$,
$\Omega^{\eta}=\eta^{a}\Omega^{l}_a$ and
$\beta=\alpha\lambda_1/\lambda_2$.

At this stage we can use the field equations (\ref{EMDFE}) which
gives
\begin{eqnarray}
&&\int_{{\cal B}} \left[|D\alpha|^2_q  + \frac{1}{2}R_{{\cal
B}}\alpha^2\right]dS \ge \int_{\cal B} \left\{ \alpha^2
|\Omega^{\eta}|^2_q
 +
 \alpha\beta |\sigma^{l}|^2_q + \alpha^2 |D\varphi|^2_q  +  2\alpha^2 V(\varphi)  \right.  \\ &&\left.+ 2\alpha\beta (l^a\nabla_{a}\varphi)^2 +
 \alpha^2 e^{-2\gamma\varphi}\left[E^2_{\perp} + B^2_{\perp}\right] + 2\alpha\beta e^{-2\gamma\varphi} (i_{l}F)_a (i_{l}F)^a +
 8\pi T_{ab}\alpha l^a (\alpha k^b + \beta l^b) \right\}dS , \nonumber
\end{eqnarray}
where $E_{\perp}=i_{k}i_{l}F$ and $B_{\perp}=i_{k}i_{l}\star F$. All
terms on the right-hand side of the above inequality are
non-negative. Indeed, for the last term we have $8\pi T_{ab}\alpha
l^a (\alpha k^b + \beta l^b)\ge 0$ since the energy-momentum tensor
of  matter satisfies the dominant energy condition. We also have
$2\alpha\beta e^{-2\gamma\varphi} (i_{l}F)_a (i_{l}F)^a\ge0$ since
the electromagnetic field satisfies the null energy condition and
$\alpha\beta\ge 0$.

Considering now the inequality for $\alpha=1$ and applying the
Gauss-Bonnet theorem\footnote{According to the Gauss-Bonnet theorem,
the Euler characteristic is given by $Euler({\cal
B})=\frac{1}{2}\int_{\cal B} R_{\cal B}dS=4\pi (1-g)$ where $g$ is
the genus of ${\cal B}$.} we find that the Euler characteristic of
${\cal B}$ satisfies
\begin{eqnarray}
Euler({\cal B})>0,
\end{eqnarray}
which shows that the topology of ${\cal B}$ is that of a
2-dimensional sphere ${S}^2$.

Discarding the  non-negative terms $\alpha\beta|\sigma^{l}|^2_q$, $
2\alpha^2 V(\varphi)$, $2\alpha\beta (l^a\nabla_{a}\varphi)^2$,
$2\alpha\beta e^{-2\gamma\varphi} (i_{l}F)_a (i_{l}F)^a $ and $8\pi
T_{ab}\alpha l^a (\alpha k^b + \beta l^b)$, we obtain

\begin{eqnarray}\label{INE1}
&&\int_{{\cal B}} \left[|D\alpha|^2_q  + \frac{1}{2}R_{{\cal
B}}\alpha^2\right]dS \ge \int_{\cal B} \alpha^2
\left\{|\Omega^{\eta}|^2_q
 + |D\varphi|^2_q  +  e^{-2\gamma\varphi}\left[E^2_{\perp} + B^2_{\perp}\right]
 \right\}dS .
\end{eqnarray}

Proceeding further, we write the induced metric on ${\cal B}$ in the
form
\begin{eqnarray}
dl^2 = e^{2C-\sigma} d\theta^2  + e^{\sigma}\sin^2\theta d\phi^2 ,
\end{eqnarray}
where $C$ is a constant. The absence of conical singularities
requires $\sigma|_{\theta=0}=\sigma|_{\theta=\pi}=C$. It is easy to
see that the area of ${\cal B}$ is given by ${\cal A}=4\pi e^{C}$.
Regarding  the 1-form $\Omega^{l}_a$, we may  use the Hodge
decomposition

\begin{eqnarray}
\Omega^{l}= *d \omega + d\varsigma ,
\end{eqnarray}
where $*$ is the Hodge dual on ${\cal B}$, and $\omega$ and
$\varsigma$ are regular axisymmetric functions on  ${\cal B}$. Then
we obtain
\begin{eqnarray}
\Omega^{\eta}= i_{\eta}*d{ \omega}
\end{eqnarray}
since $\varsigma$ is axisymmetric and
$i_{\eta}d\varsigma=\pounds_{\eta}\varsigma=0$.

We can also introduce electromagnetic potentials $\Phi$ and $\Psi$
on ${\cal B}$ defined by\footnote{We denote the Killing vector field
$\eta$ and its naturally corresponding 1-form with the same letter.}
\begin{eqnarray}
&&d\Phi = B_{\perp} * \eta, \\
&&d\Psi = e^{-2\gamma\,\varphi}E_{\perp} * \eta.
\end{eqnarray}

It turns out useful to introduce another potential $\chi$ instead of
$\omega$ which is defined by
\begin{eqnarray}
d\chi= 2 X d{\omega} - 2\Phi d\Psi + 2\Psi d\Phi ,
\end{eqnarray}
where $X=q_{ab}\eta^a\eta^b$ is the norm of the Killing field
$\eta^a$. This step is necessary in order to bring the functional
$I_{*}[X^A]$ defined below, in the same formal form as in the
stationary case.

The electric charge $Q$ and the magnetic charge $P$ associated with
${\cal B}$ are defined as follows
\begin{eqnarray}
&&Q= \frac{1}{4\pi}\int_{\cal B} e^{-2\gamma\varphi} E_{\perp} dS, \\
&&P= \frac{1}{4\pi} \int_{\cal B} B_{\perp} dS.
\end{eqnarray}
We also define the angular momentum $J$ associated with ${\cal B}$

\begin{eqnarray}
J=\frac{1}{8\pi}\int_{\cal B} \Omega^{\eta} dS +
\frac{1}{8\pi}\int_{\cal B}\left( \Phi e^{-2\gamma\varphi}E_{\perp}
 - \Psi B_{\perp}\right) dS,
\end{eqnarray}
where the first integral is the contribution of the gravitational
field, while the second integral is the contribution due to the
electromagnetic field \cite{SY}.

Using the  definitions of the potentials $\Psi$, $\Phi$, and $\chi$,
one can show that the electric charge, the magnetic charge, and the
angular momentum are given, respectively, by
\begin{eqnarray}
Q=\frac{\Psi(\pi) -\Psi(0)}{2}, \;\;  P=\frac{\Phi(\pi)
-\Phi(0)}{2}, \;\; J=\frac{\chi(\pi) -\chi(0)}{8} .
\end{eqnarray}

Since the potentials $\Psi$, $\Phi$, and $\chi$ are defined up to a
constant, without loss of generality  we put $\Psi(\pi)=-\Psi(0)=Q$,
$\Phi(\pi)=-\Phi(0)=P$, and $\chi(\pi)=-\chi(0)=4J$.

Going back to the  inequality (\ref{INE1}) and choosing
$\alpha=e^{C-\sigma/2}$,  after some algebra we obtain
\begin{eqnarray}\label{INE2}
2(C + 1)\ge \frac{1}{2\pi} \int_{{\cal B}} \left\{\sigma +
\frac{1}{4} |D\sigma|^2 + \frac{1}{4X^2}|D\chi + 2\Phi D\Psi - 2\Psi
D\Phi|^2  \right. \nonumber \\ \left.+
\frac{1}{X}e^{-2\gamma\,\varphi} |D\Phi|^2 + \frac{1}{X}
e^{2\gamma\,\varphi}|D\Psi|^2  + |D\varphi|^2\right\} dS_{0} ,
\end{eqnarray}
where the norm $|\,.\,|$  and the surface element $dS_{0}$ are with
respect to the standard usual round metric on $S^2$.  By taking into
account that ${\cal A}=4\pi e^{C}$ the above inequality is
transformed to the following inequality for the area
\begin{eqnarray}\label{INEFUN}
{\cal A}\ge 4\pi e^{(I[X^A]-2)/2} ,
\end{eqnarray}
where the functional $I[X^A]$, with
$X^A=(X,\chi,\Phi,\Psi,\varphi)$, is defined by the right-hand side
of (\ref{INE2}), i.e.
\begin{eqnarray}
I[X^A]= \frac{1}{2\pi} \int_{{\cal B}} \left\{\sigma + \frac{1}{4}
|D\sigma|^2 + \frac{1}{4X^2}|D\chi + 2\Phi D\Psi - 2\Psi D\Phi|^2
\right. \nonumber \\ \left.+ \frac{1}{X}e^{-2\gamma\,\varphi}
|D\Phi|^2 + \frac{1}{X} e^{2\gamma\,\varphi}|D\Psi|^2  +
|D\varphi|^2\right\} dS_{0} .
\end{eqnarray}

In order to bring the action into a form more suitable for  further
investigation we express $D\sigma$ by the norm of the Killing field
$\eta$ (i.e., $e^\sigma=X/\sin^2\theta$) and introduce a new
independent variable $\tau=\cos\theta$. In this way we obtain
\begin{eqnarray}
I[X^A]=  &&\int^{1}_{-1} \left\{ \frac{d}{d\tau}(\sigma\tau)  + 1 +
(1-\tau^2)\left[\frac{1}{4X^2} \left(\frac{dX}{d\tau}\right)^2 +
\frac{1}{4X^2} \left(\frac{d\chi}{d\tau} + 2\Phi\frac{d\Psi}{d\tau}
- 2\Psi\frac{d\Phi}{d\tau} \right)^2   \nonumber \right. \right.\\
&& \left.\left.  +
\frac{e^{-2\gamma\,\varphi}}{X}\left(\frac{d\Phi}{d\tau}\right)^2 +
\frac{e^{2\gamma\,\varphi}}{X}\left(\frac{d\Psi}{d\tau}\right)^2 +
\left(\frac{d\varphi}{d\tau}\right)^2\right] - \frac{1}{1-
\tau^2}\right\}d\tau .
\end{eqnarray}

At this stage we introduce the strictly positive definite
metric\footnote{It is worth mentioning  that the continuous
rotational $O(2)$ symmetry in the case of Einstein-Maxwell gravity
degenerates here to the discrete symmetry $\pm\Phi
\longleftrightarrow \pm\Psi $ and $\varphi \longleftrightarrow
-\varphi$. }
\begin{eqnarray}\label{harmetric}
 dL^2 = G_{AB}dX^A dX^B= \frac{dX^2 + \left(d\chi + 2\Phi d\Psi - 2\Psi d\Phi \right)^2 }{4X^2} + \frac {e^{-2\gamma\,\varphi}d\Phi^2
 + e^{2\gamma\,\varphi} d\Psi^2}{X} + d\varphi^2
 \end{eqnarray}
on the 5-dimensional Riemannian  manifold ${\cal N}=\{(X,\chi, \Phi,
\Psi,\varphi) \in \mr^5; X>0\}$. In terms of this metric the
functional $I[X^A]$ is written in the form
\begin{eqnarray}\label{FUN}
I[X^A]=  &&\int^{1}_{-1} \left\{ \frac{d}{d\tau}(\sigma\tau)  + 1 +
(1-\tau^2)G_{AB}\frac{dX^A}{d\tau}\frac{dX^B}{d\tau} - \frac{1}{1-
\tau^2}\right\}d\tau .
\end{eqnarray}

Let us summarize the results obtained so far in the following

\medskip
\noindent {\bf Lemma 1.} {\it Let ${\cal B}$ be a smooth, spacetime
stably outermost axisymmetric marginally outer trapped surface in a
spacetime satisfying the Einstein-Maxwell-dilaton-matter equations
(\ref{EMDFE}). If the matter energy-momentum tensor satisfies the
dominant energy condition and the dilaton potential is non-negative,
then the area of ${\cal B}$ satisfies the inequality}
\begin{eqnarray}
{\cal A}\ge 4\pi e^{(I[X^A]-2)/2},
\end{eqnarray}
{\it where the functional $I[X^A]$  is given by (\ref{FUN}) with a
metric $G_{AB}$ defined by (\ref{harmetric})}.
\medskip
\noindent

In order to put a tight lower bound for the area we should solve the
variational problem for the  minimum of the functional $I[X^A]$ with
appropriate boundary conditions if the minimum exists at all. Since
the first two terms in $I[X^A]$ are in fact boundary terms, the
minimum of $I[X^A]$ is determined by the minimum of the reduced
functional
\begin{eqnarray}
I_{\star}[X^A]=\int^{1}_{-1} \left[
(1-\tau^2)G_{AB}\frac{dX^A}{d\tau}\frac{dX^B}{d\tau} -
\frac{1}{1-\tau^2}\right] d\tau .
\end{eqnarray}

In order to perform the minimizing procedure we have to specify in
which class of functions $X^A=(X, \chi,\Phi, \Psi,\varphi)$, the
functional $I_{\star}[X^A]$ is varied. From a physical point of view
the relevant class of functions is specified by the natural
requirements $(\chi,\Phi, \Psi,\varphi)\in C^{\infty}[-1,1]$,
$\sigma=\ln\left(\frac{X}{1-\tau^2}\right) \in C^{\infty}[-1,1]$
with boundary conditions $\Phi(\tau=-1)=-\Phi(\tau=1)=P$,
$\Psi(\tau=-1)=-\Psi(\tau=1)=Q$ and
$\chi(\tau=-1)=-\chi(\tau=1)=4J$.

\medskip
\noindent {\bf Lemma 2.} {\it For a dilaton coupling parameter
satisfying  $0\le\gamma^2\le 3$, there exists a unique smooth
minimizer of the functional $I[X^A]$ (respectively $I_{\star}[X^A]$)
with the prescribed boundary conditions.}

\medskip
\noindent
{\bf Proof.} Let us consider the "truncated" functional
\begin{eqnarray}
I_{\star}[X^A][\tau_2,\tau_1]= \int^{\tau_2}_{\tau_1}\left[
(1-\tau^2)G_{AB}\frac{dX^A}{d\tau}\frac{dX^B}{d\tau}
-\frac{1}{1-\tau^2} \right] d\tau
\end{eqnarray}
with boundary conditions $X^A(\tau_1)$, $X^A(\tau_2)$ for
$-1<\tau_1<\tau_{2}<1$. By introducing the new variable
$t=\frac{1}{2}\ln\left(\frac{1+\tau}{1-\tau}\right)$, the truncated
action takes the form
\begin{eqnarray}
I_{\star}[X^A][t_2,t_1]= \int^{t_2}_{t_1}\left[
G_{AB}\frac{dX^A}{dt}\frac{dX^B}{dt} -1\right]dt,
\end{eqnarray}
which is just  a modified version of the geodesic functional in the
Riemannian space $({\cal N}, G_{AB})$. Consequently, the critical
points of our functional are geodesics in ${\cal N}$. However, it
was shown in \cite{SY} that for $0\le \gamma^2\le 3$ the Riemannian
space $({\cal N}, G_{AB})$ is simply connected, geodesically
complete, and with negative sectional curvature. Therefore, for
fixed points $X^A(t_1)$ and $X^A(t_2)$, there exists a unique
minimizing geodesic connecting these points. Hence we conclude that
the global minimizer of $I_{\star}[X^A][t_2,t_1]$ exists and is
unique for $0\le\gamma^2\le 3$. Since $({\cal N}, G_{AB})$ is
geodesically complete, the global minimizer of
$I_{\star}[X^A][t_2,t_1]$ can be extended to a global minimizer of
$I_{\star}[X^A]$ and $I[X^A]$. In more detail the proof goes as
follows. Let us put $\tau_2(\epsilon)=1-\epsilon,
\tau_1(\epsilon)=-1 + \epsilon$ (i.e.
$t_2(\epsilon)=-t_1(\epsilon)=\frac{1}{2}\ln\left(\frac{2-\epsilon}{\epsilon}\right)$
), where $\epsilon$ is a small positive number ($\epsilon>0$), and
consider the truncated functional

\begin{eqnarray}
I_{\,\epsilon}[X^A]=\int^{\tau_2(\epsilon)}_{\tau_1(\epsilon)}
\left[ \frac{d}{d\tau}(\sigma\tau) + 1\right]d\tau
  +
I_{*}[X^A][\tau_2(\epsilon),\tau_{1}(\epsilon)]= \nonumber \\
\sigma[\tau_{2}(\epsilon)]\tau_{2}(\epsilon) -
\sigma[\tau_{1}(\epsilon)]\tau_{1}(\epsilon)  + 2 (1-\epsilon) +
I_{*}[X^A][\tau_2(\epsilon),\tau_{1}(\epsilon)]
\end{eqnarray}
with boundary conditions $X^A(\tau_1(\epsilon))=X^A_1(\epsilon)$ and
$X^A(\tau_2(\epsilon))=X^A_2(\epsilon)$, and   with
$\lim_{\epsilon\to 0}X^A_1(\epsilon)=(0, 4J, P, Q,\varphi_{-})$ and
$\lim_{\epsilon\to 0}X^A_2(\epsilon)=(0, -4J, -P, -Q,\varphi_{+})$.
Here  $\varphi_{\pm}$ are defined by $\varphi_{\pm}=\varphi(\tau=\pm
1)$.

Consider now the unique minimizing geodesic $\Gamma_{\epsilon}$ in
${\cal N}$ between the points $X^A_1(\epsilon)$ and
$X^A_2(\epsilon)$. Then we have

\begin{eqnarray}\label{IEINQ}
I_{\,\epsilon}[X^A]\ge
\sigma[\tau_{2}(\epsilon)]|_{\Gamma_{\epsilon}}\,\tau_{2}(\epsilon)
- \sigma[\tau_{1}(\epsilon)]|_{\Gamma_{\epsilon}}\,
\tau_{1}(\epsilon) + 2 (1-\epsilon) +
I_{*}[X^A][\tau_2(\epsilon),\tau_{1}(\epsilon)]|_{\Gamma_{\epsilon}}
\end{eqnarray}
where the right-hand side of the above inequality is evaluated on
the geodesic $\Gamma_{\epsilon}$. Since $\lambda^2_{\epsilon}=
G_{AB}\frac{dX^{A}}{dt}\frac{dX^{B}}{dt}$ is a constant on the
geodesic $\Gamma_{\epsilon}$ we find

\begin{eqnarray}
I_{*}[X^A][\tau_2(\epsilon),\tau_{1}(\epsilon)]|_{\Gamma_{\epsilon}}=
\int^{t_2(\epsilon)}_{t_1(\epsilon)}\left[
G_{AB}\frac{dX^A}{dt}\frac{dX^B}{dt} -1\right]dt =
\left(\lambda^2_{\epsilon}-1\right)(t_2(\epsilon)- t_{1}(\epsilon)).
\end{eqnarray}

The nest step is to evaluate $\lambda_{\epsilon}$. This can be done
by evaluating $G_{AB}\frac{dX^{A}}{dt}\frac{dX^{B}}{dt}$ at the
boundary points which are in a small neighborhood of the poles
$\tau=\pm 1$. First we write $\lambda^2_{\epsilon}=
G_{AB}\frac{dX^{A}}{dt}\frac{dX^{B}}{dt}$ in the form

\begin{eqnarray}
\lambda^2_{\epsilon}= &&\frac{(1-\tau^2)^2}{4X^2}
\left(\frac{dX}{d\tau}\right)^2 + \frac{(1-\tau^2)^2}{4X^2}
\left(\frac{d\chi}{d\tau} + 2\Phi \frac{d\Psi}{d\tau} -
2\Psi\frac{d\Phi}{d\tau} \right)^2 + \nonumber\\
&& \frac{(1-\tau^2)^2}{X}e^{-2\gamma\varphi}
\left(\frac{d\Phi}{d\tau}\right)^2 +
\frac{(1-\tau^2)^2}{X}e^{2\gamma\varphi}
\left(\frac{d\Psi}{d\tau}\right)^2 + (1-\tau^2)^2
\left(\frac{d\varphi}{d\tau}\right)^2 .
\end{eqnarray}
Within the class of function that we consider, the terms associated
with $X$ and  $\varphi$ have the following behavior in a small
neighborhood of the poles $\tau=\pm 1$, namely:

\begin{eqnarray}
&&\frac{(1-\tau^2)^2}{4X^2} \left(\frac{dX}{d\tau}\right)^2= 1 +
O(\epsilon), \\
&&(1-\tau^2)^2 \left(\frac{d\varphi}{d\tau}\right)^2=O(\epsilon^2).
\end{eqnarray}

The terms associated with $\Phi$ and $\Psi$ behave, respectively,
as

\begin{eqnarray}
&& \frac{(1-\tau^2)^2}{X}e^{-2\gamma\varphi}
\left(\frac{d\Phi}{d\tau}\right)^2=O(\epsilon),  \\
&& \frac{(1-\tau^2)^2}{X}e^{2\gamma\varphi}
\left(\frac{d\Psi}{d\tau}\right)^2 =O(\epsilon).
\end{eqnarray}

In order to find the behavior of the  term associated with the
potential  $\chi$, we should notice that $\partial/\partial\chi$ is
a Killing vector for the metric $G_{AB}$, and consequently  we have
the following conservation law

\begin{eqnarray}
\frac{1}{4X^2}\left(\frac{d\chi}{dt} + 2\Phi\frac{d\Phi}{dt} - 2\Psi
\frac{d\Phi}{dt}\right)=
\frac{1-\tau^2}{4X^2}\left(\frac{d\chi}{d\tau} +
2\Phi\frac{d\Phi}{d\tau} - 2\Psi
\frac{d\Phi}{d\tau}\right)=const_{\epsilon}.
\end{eqnarray}

Using this we obtain that the  term associated with $\chi$ is equal
to $4\,const^2_{\epsilon}\,X^2$ which shows that it behaves as
$O(\epsilon^2)$. Therefore we can conclude that
$\lambda^2_{\epsilon} -1=O(\epsilon)$ which gives

\begin{eqnarray}
\lim_{\epsilon \to 0}
I_{*}[X^A][\tau_2(\epsilon),\tau_{1}(\epsilon)]|_{\Gamma_{\epsilon}}=\lim_{\epsilon
\to 0} \left(\lambda^2_{\epsilon} -1\right)(t_2(\epsilon)-
t_{1}(\epsilon))= 0.
\end{eqnarray}

In this way, from (\ref{IEINQ}) we have

\begin{eqnarray}
&&I[X^A]=\lim_{\epsilon \to 0} I_{\,\epsilon}[X^A]\ge \\
&&\lim_{\epsilon \to 0}
\left\{\sigma[\tau_{2}(\epsilon)]|_{\Gamma_{\epsilon}}\,\tau_{2}(\epsilon)
- \sigma[\tau_{1}(\epsilon)]|_{\Gamma_{\epsilon}}\,
\tau_{1}(\epsilon) + 2 (1-\epsilon) +
I_{*}[X^A][\tau_2(\epsilon),\tau_{1}(\epsilon)]|_{\Gamma_{\epsilon}}
\right\} \nonumber
\end{eqnarray}
and therefore

\begin{eqnarray}
I[X^A]\ge 2\sigma_{p} + 2
\end{eqnarray}
where $\sigma_{p}$ is the value of $\sigma(\tau)$ on the poles. This
completes the proof.

\medskip
\noindent

Even in the cases when the global minimizer of $I[X^A]$ exists,
there is another serious problem in Einstein-Maxwell-dilaton
gravity. In Einstein-Maxwell gravity the lower bound for the area is
clear from physical considerations and there is a completely
explicit solution  realizing it, namely,  the extremal Kerr-Newman
solution. So the approach is to formally prove that the area of the
extremal Kerr-Newman solution is indeed the lower bound. The
situation in Einstein-Maxwell gravity is rather different. Contrary
to the Einstein-Maxwell case where the Euler-Lagrange equations can
be solved explicitly, in Einstein-Maxwell-dilaton gravity the
corresponding Euler-Lagrange equations are not integrable for
general dilaton coupling parameter $\gamma$. So it is very difficult
to find explicitly the sharp lower bound for the area in
Einstein-Maxwell-dilaton gravity with arbitrary $\gamma$. That is
why our approach here should  be different in comparison with the
Einstein-Maxwell gravity.

\section{Area-angular momentum-charge inequality for critical dilaton coupling parameter }

The main result in this section is the  next theorem:
\medskip
\noindent

{\bf Theorem 1.}  {\it Let ${\cal B}$ be a smooth, spacetime stably
outermost axisymmetric marginally outer trapped surface in a
spacetime satisfying the Einstein-Maxwell-dilaton-matter equations
(\ref{EMDFE}) with a dilaton coupling parameter  $\gamma^2=3$. If
the matter energy-momentum tensor satisfies the dominant energy
condition and the dilaton potential is non-negative, then the area
of ${\cal B}$ satisfies the inequality}

\begin{eqnarray}
{\cal A}\ge 8\pi \sqrt{|Q^2P^2 - J^2|} ,
\end{eqnarray}
{\it where $Q$, $P$, and $J$ are the electric charge, the magnetic
charge and the angular momentum associated with ${\cal B}$,
respectively. The equality is saturated only for the extremal
stationary near horizon geometry of the $\gamma^2=3$
Einstein-Maxwell-dilaton gravity  with $V(\varphi)=0$ and
$T_{ab}=0$.}

\medskip
\noindent

{\bf Proof.}
 For the critical coupling,  $({\cal N}, G_{AB})$ is a
symmetric space with a negative sectional curvature \cite{SY}. In
fact, ${\cal N}$ is an $SL(3, R)/O(3)$ symmetric space and therefore
its metric can be written in the form
\begin{eqnarray}
dL^2= \frac{1}{8} Tr\left(M^{-1}dM M^{-1}dM\right) ,
\end{eqnarray}
where the matrix $M$ is symmetric, positive definite and $M\in
SL(3,R)$. After tedious calculations it can be shown that
\begin{eqnarray}
M=e^{\frac{2}{3}\sqrt{3} \varphi}\!\left(\!
  \begin{array}{ccc}
    X + 4\Phi^2 e^{-2\sqrt{3} \varphi} + X^{-1}(\chi + 2\Phi\Psi)^2 & 2e^{-2\sqrt{3} \varphi }\Phi + 2X^{-1}(\chi + 2\Phi\Psi)\Psi & X^{-1}(\chi + 2\Phi\Psi) \\
    2e^{-2\sqrt{3} \varphi }\Phi + 2X^{-1}(\chi + 2\Phi\Psi)\Psi  & e^{-2\sqrt{3} \varphi} + 4\Psi^2X^{-1} & 2\Psi X^{-1} \\
    X^{-1}(\chi + 2\Phi\Psi) & 2\Psi X^{-1} &  X^{-1} \\
  \end{array}\!
\right). \nonumber
\end{eqnarray}

In terms of the matrix $M$ the functional $I[X^A]$ becomes
\begin{eqnarray}\label{FUNMAT}
I[X^A]=  &&\int^{1}_{-1} \left\{ \frac{d}{d\tau}(\sigma\tau)  + 1 +
\frac{1}{8}(1-\tau^2)Tr\left(M^{-1}\frac{dM}{d\tau}\right)^2 -
\frac{1}{1- \tau^2}\right\}d\tau .
\end{eqnarray}

The Euler-Lagrange equations are then  equivalent  to the following
matrix equation
\begin{eqnarray}\label{EQM}
\frac{d}{d\tau}\left((1-\tau^2)M^{-1}\frac{dM}{d\tau}\right)=0.
\end{eqnarray}

Hence we find
\begin{eqnarray}\label{MatA}
(1-\tau^2)M^{-1}\frac{dM}{d\tau}= 2A ,
\end{eqnarray}
where $A$ is a constant matrix. From $\det M=1$ it follows that $Tr
A=0$. Integrating further we obtain

\begin{eqnarray}
M=M_{0}\exp\left(\ln\frac{1+\tau}{1-\tau}\,A\right) ,
\end{eqnarray}
where $M_{0}$  is a constant matrix with the same properties as $M$
and satisfying $A^TM_{0}=M_{0}A$. Since $M_{0}$ is positive definite
it can be written in the form $M_{0}=BB^{T}$ for some matrix $B$
with $|\det B|=1$ and this presentation is up to an orthogonal
matrix $O$ (i.e. $B\to BO$). This freedom can be used to diagonalize
the matrix $B^{T}AB^{T\;-1}$. So we can take
$B^{T}AB^{T\;-1}=diag(a_1,a_2,a_3)$ and we obtain
\begin{eqnarray}\label{MB}
M= B\left(
   \begin{array}{ccc}
     \left(\frac{1+\tau}{1-\tau}\right)^{a_1} & 0 & 0 \\
     0 &  \left(\frac{1+\tau}{1-\tau}\right)^{a_2}  & 0 \\
     0 & 0 &  \left(\frac{1+\tau}{1-\tau}\right)^{a_3}  \\
   \end{array}
 \right)B^T .
\end{eqnarray}
The eigenvalues $a_i$ can be found by comparing the singular
behavior of the left-hand and right-hand sides of (\ref{MB}) at
$\tau \to \pm 1$. Doing so we find, up to renumbering, that $a_1=0$,
$a_2=-1$, and $a_3=1$. The matrix $B$ can be found by imposing the
boundary conditions which gives
\begin{eqnarray}
B=\left(
    \begin{array}{ccc}
      -\frac{P^2Q\, e^{-\frac{1}{\sqrt{3}}(\varphi_{+}+\varphi_{-})}}{\sqrt{|P^2Q^2-J^2|}} & (2J + PQ)e^{\frac{1}{\sqrt{3}}\varphi_{-}-\frac{1}{2}\sigma_{p}}
      & (-2J + PQ) e^{\frac{1}{\sqrt{3}}\varphi_{+}-\frac{1}{2}\sigma_{p}}  \\
       -\frac{J e^{-\frac{1}{\sqrt{3}}(\varphi_{+}+\varphi_{-})}}{\sqrt{|P^2Q^2-J^2|}}& Q e^{\frac{1}{\sqrt{3}}\varphi_{-}-\frac{1}{2}\sigma_{p}} &
        - Q e^{\frac{1}{\sqrt{3}}\varphi_{+}-\frac{1}{2}\sigma_{p}} \\
      \frac{P\, e^{-\frac{1}{\sqrt{3}}(\varphi_{+}+\varphi_{-})}}{2\sqrt{|P^2Q^2-J^2|}} & \frac{1}{2} e^{\frac{1}{\sqrt{3}}\varphi_{-}-\frac{1}{2}\sigma_{p}} &
      \frac{1}{2} e^{\frac{1}{\sqrt{3}}\varphi_{+}-\frac{1}{2}\sigma_{p}} \\
    \end{array}
  \right),
\end{eqnarray}
where
\begin{eqnarray}
&&\varphi_{\pm}=\varphi(\tau=\pm1), \\
&&\sigma_{p}= \lim_{\tau\to \pm 1}
\ln\left(\frac{X}{1-\tau^2}\right)=\sigma(\tau=\pm 1).
\end{eqnarray}
Taking into account that $|\det B|=1$, we find
\begin{eqnarray}
e^{\sigma_{p}}= 2\sqrt{|P^2Q^2 - J^2|}.
\end{eqnarray}
Now we are ready to evaluate the minimum of the functional $I[X^A]$.
Substituting  (\ref{MatA}) in (\ref{FUNMAT}) we see that the last
two terms cancel each other, and we find
\begin{eqnarray}
I_{min}[X^A]= 2\sigma_{p} + 2= 2\ln\left(2\sqrt{|P^2Q^2 -
J^2|}\right) + 2.
\end{eqnarray}

Substituting further this result in (\ref{INEFUN}) we finally obtain
\begin{eqnarray}
{\cal A}\ge 8\pi \sqrt{|P^2Q^2 - J^2|}.
\end{eqnarray}
The extremal stationary near horizon  geometry is in fact defined by
equation (\ref{EQM}), by the same boundary conditions and by the
same class of functions  as those in the variational problem.
Therefore, it is clear that the equality is saturated only for the
extremal stationary near horizon geometry. This completes the proof.

\medskip
\noindent

{\bf Remark.} The case $P^2Q^2=J^2$ is formally outside of the class
of functions we consider. In the language of stationary solutions,
it corresponds to an extremal (naked) singularity with zero area.

It is interesting to note that  when $PQ=0$, but $P^2 + Q^2\ne 0$,
the lower bound of the area depends only on the angular momentum but
not on the nonzero charge in contrast with the Einstein-Maxwell
gravity.

\section{Area-angular momentum-charge inequality for dilaton coupling parameter $0\le\gamma^2\le 3$ }

As we mentioned above, finding a sharp lower bound for the area
${\cal A}$ in the case of arbitrary $\gamma$ seems to be very
difficult. However, an important estimate for the area can be found
for dilaton coupling parameter satisfying $0\le\gamma^2\le 3$. The
result is given by the following

\medskip
\noindent {\bf Theorem 2.}  {\it Let ${\cal B}$ be a smooth,
spacetime stably outermost axisymmetric marginally outer trapped
surface in a spacetime satisfying the
Einstein-Maxwell-dilaton-matter equations (\ref{EMDFE}) with a
dilaton coupling parameter $\gamma$, satisfying $0\le\gamma^2\le 3$.
If the matter energy-momentum tensor satisfies the dominant energy
condition and the dilaton potential is non-negative, then for every
$\gamma$ in the given range, the area of ${\cal B}$ satisfies the
inequality}

\begin{eqnarray}
{\cal A}\ge 8\pi \sqrt{|Q^2P^2 - J^2|} ,
\end{eqnarray}
{\it where $Q$, $P$, and $J$ are the electric charge, the magnetic
charge and the angular momentum associated with ${\cal B}$,
respectively. The equality is saturated  for the extremal stationary
near horizon geometry of the $\gamma^2=3$ Einstein-Maxwell-dilaton
gravity  with $V(\varphi)=0$ and $T_{ab}=0$.}

\medskip
\noindent

{\bf Proof.} Let us first focus on the case $0<\gamma^2 \le 3$ and
consider the metric
\begin{eqnarray}
&&d{\tilde L}^2={\tilde G}_{AB}dX^A dX^B   \\ &&= \frac{dX^2 +
\left(d\chi + 2\Phi d\Psi - 2\Psi d\Phi \right)^2 }{4X^2} + \frac
{e^{-2\gamma\,\varphi}d\Phi^2
 + e^{2\gamma\,\varphi} d\Psi^2}{X} + \frac{\gamma^2}{3}d\varphi^2
 \nonumber
\end{eqnarray}
and the associated functional
\begin{eqnarray}\label{FUNTILDE}
{\tilde I}\,[X^A]=  &&\int^{1}_{-1} \left\{
\frac{d}{d\tau}(\sigma\tau) + 1 + (1-\tau^2){\tilde
G}_{AB}\frac{dX^A}{d\tau}\frac{dX^B}{d\tau} - \frac{1}{1-
\tau^2}\right\}d\tau .
\end{eqnarray}

It is easy to see that $I[X^A]\ge {\tilde I}\,[X^A]$ and therefore
\begin{eqnarray}
{\cal A}\ge 4\pi e^{({\tilde I}\,[X^A]-2)/2}.
\end{eqnarray}
Redefining now the scalar field ${\tilde
\varphi}=\frac{\gamma}{\sqrt{3}}\varphi$, we find that the
functional ${\tilde I}\,[X^A]$ reduces to the functional $I[X^A]$
for the critical coupling $\gamma^2=3$. Hence we conclude that
\begin{eqnarray}
{\cal A}\ge 8\pi \sqrt{|Q^2P^2 - J^2|}
\end{eqnarray}
for every $\gamma$ with $0<\gamma^2\le 3$.

The case $\gamma=0$ needs a separate investigation. Fortunately, it
can be easily reduced to the pure Einstein-Maxwell case. Indeed, it
is not difficult to see that for $\gamma=0$ we have
\begin{eqnarray}
I[X^A]\ge I^{EM}[X^A] ,
\end{eqnarray}
where $I^{EM}[X^A]$ is the functional for the pure Einstein-Maxwell
gravity. In Einstein-Maxwell gravity it was proven in \cite{CJ} that
${\cal A}\ge 8\pi\sqrt{J^2 + \frac{1}{4}(Q^2 + P^2)^2}$, which gives
${\cal A}\ge 8\pi\sqrt{J^2 + \frac{1}{4}(Q^2 + P^2)^2}\ge 8\pi
\sqrt{|Q^2P^2 - J^2|}$.

Finally, it is worth noting that, as a direct consequence of Lemma
2, for every fixed $\gamma$  we obtain the following inequality

\begin{eqnarray}
{\cal A}\ge {\cal A}_{NHG}
\end{eqnarray}
where ${\cal A}_{NHG}$ is the area associated with the extremal
stationary near horizon geometry  of Einstein-Maxwell-dilaton
gravity  with $V(\varphi)=0$ and $T_{ab}=0$, for the corresponding
$\gamma$.

\section{Discussion}

In the present paper we derived area-angular momentum-charge
inequalities  for stable marginally outer trapped surfaces  in the
four dimensional Einstein-Maxwell-dilaton theory for values of the
dilaton coupling  parameter less than or equal to the critical
value. The coupling of the dilaton to the Maxwell field leads, in
general, to inequalities that can be rather different from that in
the Einstein-Maxwell gravity. Some estimates for the sector
$\gamma^2>3$ could be found if we impose the additional condition on
the dilaton potential to be convex. We leave this study for the
future.

Given the current interest in  higher dimensional gravity it is
interesting  to extend the area-angular momentum-charge inequalities
to higher dimensions. This is almost straightforward in the case of
Einstein equations \cite{H}. However, in the case of
Einstein-Maxwell and Einstein-Maxwell-dilaton gravity, the
extensions of the inequalities is difficult. The central reason
behind that is the fact that even the stationary axisymmetric
Einstein-Maxwell equations are not integrable in higher dimensions
\cite{SY1}. Nevertheless, some progress can be made, and the results
will be presented elsewhere \cite{SY2}.

\medskip
\noindent

\noindent {\bf Acknowledgements:} This work was partially supported
by the Bulgarian National Science Fund under Grant DMU-03/6, and by
Sofia University Research Fund under Grant 148/2012.


\begin{thebibliography}{99}


\bibitem{ADC} A. Acena, S. Dain and M.E. Gabach Clement, Class. Quant. Grav. {\bf 28}  105014
(2011); [arXiv:1012.2413[gr-qc]].

\bibitem{DR} S. Dain and M. Reiris. Phys. Rev. Lett. {\bf 107}, 051101 (2011); [arXiv:1102.5215[gr-qc]].

\bibitem{C} M.E. Gabach Clement, [arXiv:1102.3834[gr-qc]].

\bibitem{JRD} J. L. Jaramillo, M. Reiris and S. Dain,  Phys. Rev. {\bf D84}, 121503 (2011); [arXiv:1106.3743[gr-qc]].

\bibitem{DJR} S. Dain, J. L. Jaramillo and M. Reiris, Class. Quantum Grav. {29}, 035013 (2012); [arXiv:1109.5602[gr-qc]].

\bibitem{CJ} M.E. Gabach Clement and J.L. Jaramillo, [arXiv:1111.6248[gr-qc]].

\bibitem{S} W. Simon. Class. Quant. Grav. {\bf 29}, 062001 (2012);
[arXiv:1109.6140[gr-qc].]

\bibitem{CJR} M. E. Gabach Clement, J. L. Jaramillo and M. Reiris,
[arXiv:1207.6761[gr-qc]].

\bibitem{AHC} M. Ansorg, J. Hennig and C. Cederbaum. Gen. Rel. Grav. {\bf 43}, 1205 (2011); [arXiv:1005.3128[gr-qc]].

\bibitem{HAC} J. Hennig, M. Ansorg and C. Cederbaum. Class. Quantum Grav. {\bf 25} 162002 (2008);[arXiv:0805.4320[gr-qc]].


\bibitem{HCA} J. Hennig, C. Cederbaum and M. Ansorg, Commun. Math. Phys. {\bf 293}, 449
(2010); [arXiv:0812.2811[gr-qc]].

\bibitem{D} S. Dain. Class. Quant. Grav. {\bf 29}, 073001 (2012), [arXiv:1111.3615[gr-qc]].

\bibitem{CENS} P. Chrusciel, M. Eckstein, L. Nguyen and S. Szybka,
Class. Quant. Grav. {\bf 28}, 245017 (2011).

\bibitem{Mars} M. Mars, Class. Quant. Grav. {\bf 29}, 145019 (2012).

\bibitem{J} J. Jaramillo, Class. Quant. Grav. {\bf 29}, 177001
(2012).

\bibitem{Gibbons} G.~ Gibbons~ and~ K.~ Maeda,~ Nucl.~ Phys.~{ \bf B298}, 741
(1988).

\bibitem{Garfinkle} D.~ Garfinkle, G.~Horowitz~ and~ A.~ Strominger, Phys.~ Rev.~{ \bf D43}, 3140 (1991);
{\bf D4}5, 3888, 1992 (E).

\bibitem{Maison} D.~ Maison, Gen.~ Rel.~ and~ Grav. {\bf 10}, 717 (1979).

\bibitem{Hojman} S.~ Hojman, M.~ Rosenbaum~ and~M. ~Ryan, Phys.~Rev.~{\bf D17}, 3141 (1978).

\bibitem{SY}  S.~Yazadjiev, Phys. Rev. {\bf D82}, 124050 (2010); [arXiv:1009.2442[hep-th]]

\bibitem{UAMS}  A.~ Masood-ul-Alam,  Class.~Quant.~Grav.~{\bf 10}, 2649 (1993);
M.~Mars~and~W.~Simon, Adv.~Theor. ~Math.~Phys.~ {\bf 6}, 279 (2003).


\bibitem{Kunz} B.~Kleihaus, J.~ Kunz~and~F. Navarro-Lerida, Phys.~Rev.~ {\bf D69}, 081501 (2004); [arXiv:0309082[gr-qc]].

\bibitem{H} S. Hayward, Phys. Rev. {\bf D49}, 6467 (1994).

\bibitem{AMS}  L. Andersson, M. Mars and W. Simon. Phys. Rev. Lett. {\bf 95}, 111102
(2005);[arXiv:0506013[gr-qc].]

\bibitem{AMS1} L. Andersson, M. Mars and W. Simon. Adv. Theor. Math. Phys., {\bf 12(4)}, 853
(2008); [arXiv:0704.2889[gr-qc]].

\bibitem{BF} I. Booth and S. Fairhurst, Phys. Rev. {\bf D77}, 084005 (2008); [arXiv:0708.2209[gr-qc]].

\bibitem{H} S. Hollands, Class. Quant. Grav. {\bf 29}, 065006 (2012); [arxiv:1110.5814[gr-qc]].

\bibitem{SY1} S. Yazadjiev,  JHEP {\bf 1106}, 083 (2011); [arXiv:1104.0378
[hep-th]].

\bibitem{SY2} S. Yazadjiev, In preparation












\end{thebibliography}
\end{document}